# Magneto-Coriolis Waves in a spherical Couette flow experiment

Denys Schmitt,[a] P. Cardin, P. La Rizza and H.-C. Nataf
ISTerre, Université de Grenoble 1 - CNRS, F-38041 Grenoble, France

February 2012

**Abstract**

The dynamics of fluctuations in a fast rotating spherical Couette flow experiment in the presence of a strong dipolar magnetic field is investigated in detail, through a thorough analysis of the experimental data as well as a numerical study. Fluctuations within the conducting fluid (liquid sodium) are characterized by the presence of several oscillation modes, identified as magneto-Coriolis (MC) modes, with definite symmetry and azimuthal number. A numerical simulation provides eigensolutions which exhibit oscillation frequencies and magnetic signature comparable to the observation. The main characteristics of these hydromagnetic modes is that the magnetic contribution has a fundamental influence on the dynamical properties through the Lorentz forces, although its importance remains weak in an energetical point of view. Another specificity is that the Lorentz forces are confined near the inner sphere where the dipolar magnetic field is the strongest, while the Coriolis forces are concentrated in the outer fluid volume close to the outer sphere.

---

[a] Author to whom correspondence should be addressed. Electronic mail: Denys.Schmitt@obs.ujf-grenoble.fr

# I. INTRODUCTION

Magnetohydrodynamics (MHD) laboratory experiments are of great importance to understand the dynamics of the Earth's outer core, in relation to the generation of its magnetic field [1]. Although they do not reproduce the full complexity of the core, they can mimic several particular aspects, providing quantitative estimates of specific properties of the flow. One of the major benefits of laboratory experiments is their ability to include nonlinear effects and their associated instabilities that naturally accompany large-amplitude forcing in fluids. Extreme cases include experimental dynamos, where large magnetic Reynolds numbers were reached owing to the use of a highly conductive fluid, namely sodium, large velocities and an appropriate geometry, allowing to obtain self-sustained magnetic fields [2, 3, 4]. More modest installations often involve rapid rotation, because this property is common in geophysical and astrophysical objects, and the use of an imposed magnetic field, which allows to observe the magnetic signature of the turbulent flow or, if the magnetic field is strong enough, to substantially influence the flow itself.

In this context, the "Derviche Tourneur Sodium" (DTS) experiment has been designed to explore a magnetostrophic dynamical regime similar to the one supposed to occur within the Earth's outer core, i.e. where both magnetic (Lorentz) and Coriolis forces are in balance [5]. It consists in a rotating spherical Couette flow device where liquid sodium, used as the conducting fluid, is immersed in a strong imposed dipolar magnetic field [6]. Besides the superrotation phenomenon, observed in the vicinity of the inner sphere [7], hydromagnetic waves have been also evidenced within the fluid as soon as a differential rotation between inner and outer spheres is present [8]. Several types of waves, all retrograde with regard to the mean fluid angular velocity, have been identified depending on the relative rotation rates of both spheres, and their main characteristics have been determined. However, the exact nature of these waves remained ambiguous, although the close proximity of the magnetic and inertial time scales led them to be identified as magneto-inertial (or magneto-Coriolis) modes. To go deeper into the understanding of these modes, new extensive measurements of the magnetic field induced at the surface of the outer sphere have been performed in the case where the outer sphere is kept at rest, allowing us to extract their symmetrized components. In a parallel and complementary way, a new numerical code has been developed to calculate the properties of hydromagnetic waves in DTS configuration [9]. These new experimental and numerical results are analyzed in the present work. Section II focuses on the experimental part, while the results of numerical simulation are described in Section III. The last Section is devoted to the discussion.

## II. EXPERIMENTAL

### A. Experimental details

Details of the experimental set-up of DTS have been widely described previously [see 6-8] and are summarized here below. Forty liters of liquid sodium are contained between a 7.4 cm radius inner sphere and a 21 cm radius outer sphere. The copper inner sphere contains several magnetized bricks producing a nearly dipolar magnetic field within the sodium, their magnitude ranging from 0.008 T to 0.345 T throughout the volume of fluid. This imposed magnetic dipole is aligned with the rotation axis of the device. The inner and outer spheres can be independently set in rotation around a common axis, their rotation rates being servo-controlled to remain constant. Note however that all the results presented here below have been obtained with the outer sphere kept at rest. The only forcing then originates from the rotation frequency $\Delta f$ of the inner sphere, which induces strong shears within the flow [7].

The magnetic field outside the spherical fluid shell is measured by a set of Giant Magneto-Resistance (GMR) sensors arranged on several soft strips fixed on the outer sphere along a meridian. Twenty positions are instrumented between −55° and +55° of latitude. At each location, the three components of magnetic field are measured by three distinct sensors, one aligned perpendicular to the sphere, and the other two tangent to the sphere and at ±45° of the meridian, respectively. This configuration allows us to ensure that all the sensors are well polarized by the background dipolar magnetic field, in order to be used within a range where their sensitivity is nearly constant (about 40 mV/mT). Actually, for some extreme locations (high latitudes), this is not the case because the polarizing field is too large, and the corresponding radial measurements have to be corrected according to the typical response curve of the GMR sensors before a quantitative analysis. Moreover it should be emphasized that the GMR probes are sensitive to the magnitude, not to the sign of the magnetic field. As we are concerned with small fluctuations, care will have to be taken with this property since a positive value of the induced field will lead to an increase or a decrease of the signal according to the positive or negative sign of the background field, respectively. Finally, while the sensors perpendicular to the sphere provide directly the radial magnetic field $b_r$, the measurements arising from the other two sensors have to be combined (sum and difference) in order to obtain the meridional $b_\theta$ and azimuthal $b_\varphi$ components of the field, respectively.

As it will be detailed in Section III, and as far as the hydromagnetic waves in DTS are concerned, symmetry considerations with regard to the equator can be applied to their two components, namely the fluid velocity $\boldsymbol{u}$ and the magnetic field $\boldsymbol{b}$. More precisely, the vector fields $\boldsymbol{u}$ and $\boldsymbol{b}$ associated with a particular wave are either equatorially symmetric (*ES*) or antisymmetric (*EA*). Accordingly, each of their components has a specific parity by equatorial symmetry, i.e. $(v_r, v_\theta, v_\varphi)$ will transform to $(v_r, -v_\theta, v_\varphi)$ for a *ES* vector field $\boldsymbol{v} = \boldsymbol{u}$ or $\boldsymbol{b}$, or to $(-v_r, v_\theta, -v_\varphi)$ for a *EA* vector field. Therefore, the appropriate combinations $b_{even} = \frac{1}{2} [b(\theta) + b(\pi-\theta)]$ and $b_{odd} = \frac{1}{2} (b(\theta) - b(\pi-\theta)]$ are worth being calculated from measurements performed at colatitudes $\theta$ and $\pi-\theta$, symmetric with regard to the equator, in order to obtain the symmetrized magnetic components $b_{r,sym}$, $b_{\theta,sym}$ and $b_{\varphi,sym}$.

## B. Experimental results

## 1. Radial magnetic field

In a typical sequence of measurement, the outer sphere being kept at rest, the inner sphere is maintained at a constant rotation rate for about five minutes before changing to another rotation rate. During each successive plateau, the fluid also rotates at a stationary rate. As far as the dynamical properties are concerned, these long plateaus allow to accumulate the fluctuation measurements during a long period of time, in order to increase the *signal to noise* ratio of the Power Spectral Density (PSD). As an example, the PSD of the radial induced magnetic field corresponding to $\Delta f$ = 21.6 Hz is shown in Fig. 1, for the ten sensors of the northern hemisphere (the corresponding mean angular velocity of the flow is 15.4 Hz). A succession of broad bumps at different frequencies is observed, in agreement with the previous study [8]. They correspond to modes propagating azimuthally with successive wavenumbers $m$ = 1, 2... up to 7 (modes with a higher wavenumber are hardly visible). Moreover, the evolution of their relative amplitude can be followed as a function of latitude: in particular, the $m$ = 1 mode around 10 Hz becomes much stronger and dominant when approaching the equator while the relative amplitude of $m$ = 2 (~18 Hz) and 3 (~26 Hz) modes reverses between low and high latitudes (see below for more quantitative details).

To go further in the analysis of these modes, a symmetrization of the radial component has been performed by considering the sum and difference of raw data measured at latitudes symmetric with regard to the equator. So the even $b_{r,\text{even}}$ and odd $b_{r,\text{odd}}$ parts of the radial magnetic field have been extracted. Their corresponding PSD's are shown in Fig. 2, with a linear scale in order to better appreciate the latitudinal variation of their amplitude. It appears that the even and odd radial components exhibit a very different behaviour. The even part is dominated by modes with $m$ = 1 (within the range [5-12] Hz) and $m$ = 3 ([22-30] Hz) azimuthal numbers while $m$ = 2 modes ([15-21] Hz) are dominant for the odd part. In both cases, other minor peaks are present. It turns out that each major or minor peak has a specific location in frequency, constant throughout the latitude. Besides, some broad peaks show a substructure. For example, the *m* = 2 odd component includes two sub-components (see arrows in Fig. 2b): a first, weak sub-component $b_{r,odd}^{a}$ is present mainly below ~30° latitude around 15.5 Hz, while the second one $b_{r,odd}^{b}$ occurs at the frequency 19 Hz and is the strongest between 20° and 40° latitude. This feature may explain why the bumps observed on the *total* PSD seem sometimes to shift along the meridian (see Fig. 1): while the frequency of each sub-component remains constant according to the latitude, the change of relative magnitude of both sub-components may give illusion of a frequency drift along the meridian. Note also that all high latitude modes are strongly attenuated, a part of this reduction being due to the saturation of the GMR sensors in the polarizing background magnetic field. The thorough analysis of the data described here emphasizes the wealth of the dynamics present in the DTS flow, since no less than 14 different modes have been identified for $m \leq 4$.

It should be noticed that the symmetrization described above is valid only if the modes are present in the whole fluid volume. This has been indeed verified by performing correlations between data obtained at latitudes symmetric with regard to the equator, after the data have been filtered using a sliding window band-pass filter. For a given pair of latitudes (e.g. ±35° in Fig. 3), intense spots emerge at the same frequencies as the dominant PSD peaks, with a color (at zero time lag) related to the parity of the corresponding modes : red for an even mode, blue for an odd one. The same behaviour occurs for all the other latitudes. At frequencies where both parities are equally present, e.g. around 16 Hz, both contributions cancel out and therefore do not appear. The level of coherence between dominant modes at symmetric latitudes is directly given by the color scale, as the cross-correlation has been normalized for each frequency so that the autocovariance at zero lag is identically unity. This is true even for very weak modes, as those seen at 6 and 9 Hz for odd parity. The coherence thus ranges between 50 and 80% according to the mode under consideration. The existence and the strength of these intense spots establish the good coherence of the associated modes across the equator, throughout a large fraction of the full volume.

A last important result can be obtained about the time shift of the data along a meridian, i.e. a possible spiralization of the modes. To accurately estimate a possible time shift of the modes according to the latitude, a correlation of the data has been performed between a given latitude (pivot) and all the others in the opposite hemisphere, after the signals have been filtered around the frequency of the mode under consideration (band-pass filter). Note that the choice of the opposite hemisphere is only a way to emphasize the different behaviour of even and odd parity components. As an example, the correlations for the two even and the two odd modes are reported in Fig. 4 for $m = 1$. In these figures, the signal measured at 35° latitude has been taken as the pivot, and each cross-correlation has been normalized so that the autocovariance at zero lag is identically unity. This normalization allows us to follow more easily the spiraling of the modes even at latitudes where the signal is weak. It appears that the $b_{r,\text{even}}^{a}$ and $b_{r,\text{odd}}^{b}$ modes are strongly spiraled in *westward* direction from the equator, i.e. the signal around the equator is *early* compared to those at high latitudes (the fluid and the waves are moving *eastward*). The maximum time shift reaches ~55 ms (~60 ms) for $b_{r,\text{even}}^{a}$ ($b_{r,\text{odd}}^{b}$), i.e. about 40% (70%) of the wave period, respectively. For the other two modes, the maximum time shift is noticeably smaller, remaining below ~10% of the period, and the spiraling is not regular but quite significant. This behaviour is reminiscent of the convective Busse's columnar rolls, where the spiralization results from the combined effects of strong Coriolis forces, spherical boundaries, and moderate viscous dissipation [10, 11], or of the spiralization of the Rossby waves observed in the destabilization of the Stewartson layers [12]. This will be discussed in Section IV. For modes with higher azimutal numbers $m$ (2, 3...), the spiralization is less pronounced but a small variation nevertheless subsists along the meridian.

## 2. Meridional and azimuthal magnetic field

PSD of the symmetrized meridional and azimuthal components of the fluctuating magnetic field are shown in Fig. 5 and 6, respectively. The behaviour noticeably differs from that of the radial components. For the meridional component, both even and odd parts are dominated by the $m = 1$ and $2$ azimuthal numbers, mainly at low latitudes, while $m = 3$ (even part) and $m = 2$ (odd part) contributions are predominant for the azimuthal component. Moreover, the amplitude of fluctuations remains important even at the highest latitudes investigated, where the GMR sensors are not saturated by the imposed magnetic field. Again, the broad peaks seem to be resolved into two sub-components, as for example the $m = 2$ component of $b_{\theta,\text{even}}$, at about 16 and 19.5 Hz, or the $m = 3$ component of $b_{\varphi,\text{odd}}$, around 25 and 29 Hz. A total of 14 different modes may then be identified for $m \leq 4$, the $m = 4$ modes being always very weak.

As explained in the previous section, the latitudinal evolution of the tangential field components $b_\theta$ and $b_\varphi$ can be also followed through the correlations between a given latitude (pivot) and the other ones. Here, due to the large number of simultaneous measurements necessary to extract the even and odd tangential components, the measurements could not be performed along the full meridian at the same time, but separately at high then low latitudes. A prominent feature appears to be the phase inversions occurring around 15° latitude for the $b_{\theta,\text{even}}$ component (frequency range 14-16 Hz), and around 40° latitude for the same component around 20 Hz, while their spiralization remains weak (not shown). This phase inversion means that the corresponding magnetic field component changes its sign at a given latitude. This constitutes a strong constraint on the geometry of the fluctuating mode under consideration. A last important characteristic of these variations is the absence of noticeable spiralization for all the azimuthal components.

## C. Reconstruction of the magnetic potential

The magnetic field components shown in the previous sections have been measured immediately outside the outer sphere of DTS experiment, i.e. in an insulating region. As a consequence, these three components are not independent from each other since they derive from the same magnetic potential $\phi$. For a given azimuthal number $m$, they can be expanded as:

$$\begin{cases} b_r(\theta,\varphi,t) = \sum_{l \geq m}(l+1)r_b^{-l-2}\phi_l^m Y_l^m(\theta,\varphi)e^{-i\omega t} + cc \\ b_\theta(\theta,\varphi,t) = -\sum_{l \geq m}r_b^{-l-2}\phi_l^m \frac{\partial}{\partial \theta}Y_l^m(\theta,\varphi)e^{-i\omega t} + cc \\ b_\varphi(\theta,\varphi,t) = \sum_{l \geq m}\frac{im}{\sin(\theta)}r_b^{-l-2}\phi_l^m Y_l^m(\theta,\varphi)e^{-i\omega t} + cc \end{cases} \quad (1)$$

where $r_b$ is the radial position of the magnetic probes, $\omega$ the angular frequency of the mode under consideration, $\phi_l^m$ the successive (complex) coefficients of the magnetic potential and *cc* stands for *complex conjugate*.

It has been shown that the modes evidenced in DTS are far from being permanent, exhibiting a mean lifetime of ~40 rotation periods, i.e. a few seconds [8]. Therefore, refining the coefficients $\phi_l^m$ from the time dependence of the three raw components is not appropriate to deduce the spatial characteristics of the modes. Following the time-evolution of the complete magnetic potential would require a global coverage of the sphere with magnetometers, while we can use a single meridian only because we work in the spectral domain. We therefore consider the correlations between various components at different latitudes, because these quantities provide pertinent information on the relative time and spatial variation of these components, *in average*. In particular, if the envelope of the correlation functions gives an estimation of the lifetime of the modes, the position of their first maximum (amplitude and phase) is directly related to their morphology. These data can

- 10 -then be used to refine the coefficients $\phi_l^m$ by comparing the maximum of the calculated correlation functions between various components in Eq. 1 with the experimental values. If the field components in Eq. (1) are rewritten as:

$$b_q(\theta,\varphi,t) = b_q(\theta)e^{i\phi_q} e^{i(m\varphi-\omega t)} + cc \qquad (2)$$

where $q = r$, $\theta$ or $\varphi$, $b_q(\theta)$ is real and $\phi_q$ is the associated phase, the correlation function between components $p$ and $q$ takes the form:

$$C_{pq}(\theta,\theta',\tau) = 2b_p(\theta)b_q(\theta')\cos(\phi_p - \phi_q - \omega\tau) \qquad (3)$$

and its maximum occurs for $\tau = (\phi_p - \phi_q)/\omega$.

To refine the coefficients $\phi_l^m$, the following two-step procedure has been used. First, the field components $(b_q(\theta), \phi_q)$ at various colatitudes $\theta$ have been deduced from a set of experimental correlation functions (Eq. 3), one particular component being taken as the reference, namely $b_r$ at 25° latitude (associated phase $\phi_r = 0$). Then the coefficients $\phi_l^m$ have been refined by using Eqs. (1) and (2) in a linear least-square minimization (the maximum value for $l$ has been taken as low as possible according to the reduced number of experimental data). This procedure has been performed for the three most intense modes shown in the previous sections, namely the $S_{1-+}^b$, the $S_{2+-}^b$ and the $S_{3-+}^a$ modes, centered around 11.5, 20 and 27 Hz, respectively. In such a $S_{m+-}^k$ notation, $k = a$, b... corresponds to the successive modes for a given azimuthal wavenumber $m$, and the sign + (-) refers to a *ES* (*EA*) vector field: for the $S_{m+-}^k$ mode for example, the velocity component is *ES* and the magnetic field *EA*.

For the $S_{1-+}^b$ mode ($m = 1$), six $\phi_l^m$ coefficients ($l = 2p-1$, $1 \leq p \leq 6$) are necessary to satisfactorily describe the experimental data (see Fig. 7). The dominant coefficient is the first one ($l = 1$), corresponding to the largest scale, while the next four ones are one order of magnitude weaker but remain necessary to explain the spiralization of the data. This latter is particularly pronounced for the $b_\theta$ component. Moreover, a secondary maximum is also visible at high latitude (~80°) for $b_r$, but its magnitude is not well constrained because of the lack of experimental data in this region. The corresponding magnetic potential reconstructed at the surface of the outer sphere is shown in Fig. 8: a large pattern is located around a latitude of ~25°, and a weak but clear spiralization is visible at low latitude. Note that considering less coefficients in the refinement leads to a worse agreement with no spiralization or exaggerated secondary maximum.

The same procedure as above has been followed for $m = 2$ and 3, namely for the $S^b_{2+-}$ and $S^a_{3-+}$ modes. For the $S^b_{2+-}$ mode, the refinement of the magnetic potential requires at least four coefficients to obtain a satisfactory agreement. The corresponding magnetic potential is more localized (around 30°) than for $m = 1$ - note that the magnetic field is *EA* here - and the spiralization is weaker (Fig. 8). Again, the first coefficient is dominant for this mode, and it corresponds to $l = 3$ for the sake of symmetry. Note that the $b_r$ and $b_\theta$ components are in-phase at low latitude, but a change of sign of $b_\theta$ occurs clearly at ~40° latitude (not shown). Finally, for $m = 3$, five coefficients are necessary to provide a good description of the data. Here also, the first coefficient (i.e. $\phi^3_3$) is dominant, and the spiralization of $\theta$-component is very pronounced below a latitude of ~40°.

## III. NUMERICAL

### A. Numerical details

The details of the model used to calculate the oscillating modes propagating in a spherical shell of an incompressible conducting fluid within an external magnetic field have been widely described previously [9]. They are summarized here below, the notation being adapted to the present situation. The fluid is enclosed between an outer and an inner sphere of radii $a$ and $\gamma a$, respectively. The outer sphere is kept at rest, while the inner sphere rotation rate is $\Delta\Omega$. The linearized equations describing the oscillating modes (MC waves) are written as:

$$\begin{cases} \dfrac{\partial \mathbf{u}}{\partial t} = -\nabla p - (\mathbf{U_0}\cdot\nabla)\mathbf{u} - (\mathbf{u}\cdot\nabla)\mathbf{U_0} + Le^2((\nabla\times\mathbf{B_0})\times\mathbf{b} + (\nabla\times\mathbf{b})\times\mathbf{B_0}) + E\Delta\mathbf{u}, \\ \dfrac{\partial \mathbf{b}}{\partial t} = \nabla\times(\mathbf{U_0}\times\mathbf{b}) + \nabla\times(\mathbf{u}\times\mathbf{B_0}) + E_m\Delta\mathbf{b}, \\ \nabla\cdot\mathbf{u} = 0, \quad \nabla\cdot\mathbf{b} = 0 \end{cases} \quad (4)$$

where $p$ is the pressure, $\mathbf{u}$ and $\mathbf{b}$ represent the velocity and magnetic field components of the MC-wave, respectively, while $\mathbf{U_0}$ and $\mathbf{B_0}$ are the velocity and magnetic field describing the background state, assumed to be known (analytically or numerically) and always axisymmetric. Here, the reference frame is taken as the rest (laboratory) frame, because the fluid is not rotating at the same rate through the whole volume. In the motion and induction equations, all the quantities are dimensionless, the characteristic scales being $a$ for the length, $a\gamma\Delta\Omega$ for the velocity and $B_0^{eq}$, the magnitude of the field at the outer sphere equator, for the magnetic field. The corresponding dimensionless parameters involved in both equations are then the Lehnert number $Le = B_0^{eq}/(a\gamma\Delta\Omega\sqrt{\rho\mu})$, the Ekman number $E = \nu/(a^2\gamma\Delta\Omega)$ and the magnetic Ekman number $E_m = \eta/(a^2\gamma\Delta\Omega)$. Other related dimensionless parameters are the magnetic Reynolds number $R_m = 1/E_m$ and the Elsasser number $\Lambda = Le^2 R_m$.

The divergence-free property allows us to separate each vector field into a poloidal $v_p$ and a toroidal $v_t$ part, and their general expression is assumed to be $v = v(r,\theta)\exp[i(m\varphi - \tilde{\omega} t)]$, with $m$ the azimuthal wavenumber and $\tilde{\omega}$ the complex frequency. Note that $m = 0$ and $\tilde{\omega} = 0$ for the background vector fields $\mathbf{U_0}$ and $\mathbf{B_0}$. Moreover, owing to the spherical symmetry, the poloidal and toroidal scalars are expanded in spherical harmonics. These scalar components of the eigenmodes (**u**, **b**) are then obtained by solving the magneto-inertial equations (Eqs. 4), providing also the corresponding complex eigenvalue ($\omega$ + i$\lambda$), where $\omega$ is the real eigenfrequency and $\lambda$ the attenuation rate. The boundary conditions used in the present study are the no-slip condition for the velocity on both spheres, and the insulating (perfectly conducting) condition for the magnetic field on the outer (inner) sphere, respectively. Other symmetry considerations imply that i) there is no coupling different *m*'s for the eigenmodes, and ii) two types of vector fields **v** may be considered, whether they are symmetric (*ES*) or antisymmetric (*EA*) with respect to the equator. Therefore, and due to the symmetry of **U$_0$** (*ES*) and **B$_0$** (*EA*) in the *DTS* experiment, only two groups of symmetry are allowed for the eigenmodes, i.e. G2 and G3 groups (see 9). For the G2 group, **u** is *ES* and **b** *EA*, while the reverse is true for the G3 group.

As far as the basic state (**U$_0$**, **B$_0$**) is concerned, it has been obtained from an independent numerical simulation by using the non-linear 3D PARODY code [13]. The dimensionless parameters $Le$ = 0.25, $E$ = 2.5 $10^{-4}$ and $E_m$ = 0.25 have been used, because they ensure a magnetostrophic equilibrium for the fluid, in which Lorentz and Coriolis forces are in balance (Elsasser number $\Lambda$ = $Le^2/E_m$= 0.25). They are thought to reasonably describe the experimental results, as far as the mean axisymmetric properties are concerned [14, 8, 7]. Meridional section of the angular velocity $u_\phi$/(r sin$\theta$) of the background vector field **U$_0$** is shown in Fig. 9a; in particular, the superrotation phenomenon is well reproduced near the equator and close to the inner sphere. In the remaining volume, the fluid is almost in solid body rotation, except near the outer sphere around the equator where it takes a geostrophic character; this is also consistent with the observation in DTS. Note also that the induced toroidal magnetic component $B_\phi$ is weak in this calculation and slightly underestimated compared to the experiment. A second background state has been occasionally used to check the influence of the basic state on the calculated eigenfrequencies: it arises from the XSHELLS code, developed in order to study magnetic spherical Couette flows [15]. In this simulation, the boundary conditions for the magnetic field differ from the PARODY calculation, since the true conductivity has been considered for both inner and outer spheres. Then, the parameters $Le = 0.388$, $E = 3.83 \ 10^{-4}$ and $E_m = 0.383$ have been used. The corresponding axisymmetric flow is quite similar to the PARODY results (see Fig. 9b), but the details of small scale features may differ. As far as the induced magnetic field is concerned, it turns out that it also remains weak with respect to the imposed dipolar field. Note that for both background states, the three parameters used in the present calculation of MC-modes have been taken as those of the corresponding background state, for sake of consistency.

## B. Numerical results

A first important point to be addressed is the comparison between the experimental and calculated MC-mode frequencies. The experimental frequencies $f_{exp}$ of the modes described in the previous section are gathered in the Table. They all correspond to the same forcing, namely a rotation frequency of the inner sphere $\Delta f = 21.5$ Hz. Their experimental dispersion is about $\pm 1$ Hz (see Figs. 2, 5, 6). As the mode frequencies exhibit a nearly linear dependence on the forcing rate [8], the normalized frequencies $f_{exp}/\Delta f$ appear to be characteristic of each mode. As far as the numerical simulation is concerned, the mode calculation is linear, therefore the ratio $\omega_{calc}/\Delta\Omega^*$ will be taken as the characteristic value to be compared to $f_{exp}/\Delta f$. For the two background states (PARODY and XSHELLS), the dimensionless inner sphere rotation rate is $\Delta\Omega^* = 2.86$. For sake of conciseness, only the three most intense experimental modes will be considered here below, namely the $S^b_{1-+}$, the $S^b_{2+-}$ and the $S^a_{3-+}$ modes.

*m* = 1 results

For the azimuthal wavenumber $m = 1$, the dominant experimental MC-mode is $S^b_{1-+}$ (G3 symmetry), its normalized frequency is $f_{exp}/\Delta f \approx 0.51$. If the calculations are restricted to eigensolutions with a weak attenuation rate, only few solutions exist in the range $\omega_{calc} = [0, 3]$, i.e. around $0.51\ \Delta\Omega^*$ (see Fig. 10). Among them, one solution is found with a normalized rotation rate $\omega_{calc}/\Delta\Omega^* = 0.51$ very close to the experimental ratio, but its corresponding spatial characteristics noticeably differ from the observation, showing a similar pattern but a wrong spiralization (not shown). Another solution with $\omega_{calc} = 1.11$ ($\omega_{calc}/\Delta\Omega^* = 0.39$) might better correspond to the observed $S^b_{1-+}$ mode and is shown in Fig. 11. A remarkable feature of this mode lies in its large scale characteristics, in particular as far as its magnetic component is concerned. In comparison, the velocity exhibits intense additional shear zones, in particular near the pole axis and in the vicinity of the outer sphere. Note that the mode velocity remains weak in the superrotation region, while the magnetic field components seem to ignore this region (they do not follow its contour). This different behaviour between velocity and magnetic field could be related to the low magnetic Prandtl number used in the calculations, namely $P_m = 10^{-3}$, which implies that the magnetic features are smoothed out by the relatively large magnetic Ekman number. It is worth noting that the choice of meridian ($\varphi = 10°$ in Fig. 11) is somewhat arbitrary, the velocity and magnetic components evolving as a function of the longitude ; the corresponding figures can be located with respect to the dominant magnetic patterns present on the outer sphere (Fig. 11b). A second important feature of the calculated MC-mode is the slight spiralization of the magnetic field components just on the outer sphere, as seen on the magnetic potential (Fig. 11b). The phase shift between the equator and a latitude of ~35° reaches about 6°, a value four times smaller than the observed one (24°), but with a spiralization in the right direction (compare with Fig. 8a). Moreover, the evolution of the potential amplitude as a function of latitude is in good agreement with the experiment, the mode being localized near the equator, with a maximum around 15°. These variations are exemplified by following the amplitude and azimuthal position of the maximum of potential as a function of latitude (see Fig. 12).

Considering the other possible MC-modes, it can be noticed that one instable mode (positive attenuation rate) is found to occur at the normalized angular frequency 0.82. Such a mode is expected to grow within the fluid and then to saturate due to non-linear terms. However, no mode has been found experimentally around this frequency, which is noticeably higher that the observed ones. Focusing on the damped MC-modes, it turns out that their density in the ($\lambda$, $\omega$) plane remains moderate, when compared to the density of the solutions obtained by removing the Lorentz forces (i.e. $Le = 0$, see Fig. 10). Actually, without magnetic forces, the solutions are inertial-type modes, with spatial features dramatically distorted compared to the canonical inertial modes [16] because of the presence of strong shears within the background velocity field. Moreover, the magnetic diffusivity plays no role in these modes, the attenuation of which only arises from the viscosity through the Ekman number. As soon as the Lorentz forces are involved ($Le \neq 0$), the influence of the magnetic Ekman number grows, explaining the strong increase of the attenuation ($\lambda$ values more negative) and consequently the lowering of the mode density.

The influence of the $m = 0$ background magnetic field induced by the mean flow on the modes has been checked. It turns out that the solutions obtained by including this induced field are very close to those calculated with dipolar field only (see Fig. 10). This is in agreement with the weakness of the induced field compared to the main dipolar field, which remains a dominant ingredient for determining the mode properties. The influence of the boundary conditions has also been checked, by considering an insulating inner sphere instead of a perfectly conducting one. The results are extremely close to each other, as shown in Fig. 10. This conclusion is drastically different from the calculation of the background state itself, where the presence of a conducting inner sphere is the key to obtain a vigorous coupling with the fluid and to get the super-rotation phenomenon [7, 17].

Finally, the influence of the background state has been investigated. In Fig. 10, the solutions obtained with the second background state (XSHELLS code) are reported. They are quite different from the PARODY solutions, while the background mean velocity seems to be rather similar (see Fig. 9). It follows that actually the fine details of the shear zones have a fundamental importance on the final mode frequencies. Nevertheless, the mode density in the ($\lambda$, $\omega$) plane remains comparable, and there exists also one instable mode in the investigated frequency window.

*m* = 2 results

For $m = 2$, the dominant experimental MC-mode is $S^b_{2+-}$ (symmetry G2), its normalized frequency is $f_{exp}/\Delta f \approx 0.88$. Some calculated eigensolutions ($\omega$, $\lambda$) are reported in Fig. 13 in the region of interest, for different parameters. As quoted above for $m = 1$, the solutions are not very dense in the ($\lambda$, $\omega$) plane for the PARODY parameters. Among the solutions, the mode with $\omega_{calc}/\Delta\Omega^* = 0.94$ appears to well describe the main features of the experimental $S^b_{2+-}$ mode: its calculated frequency is very close, and its attenuation rate is the weakest at this frequency. Note that the location of this frequency may appear surprising, since all the solutions in the vicinity exhibit a lower (i.e. less negative) attenuation rate. It can be emphasized that the behaviour of the magnetic potential at the outer sphere surface is similar to the experimental one (Fig. 14): in particular, its maximum occurs at about 35°, and its spiralization is close to the experimental one (Fig. 12).

In order to appreciate the influence of the forcing $\Delta\Omega$ on the MC-modes, calculations have been performed for several set of parameters between the main configuration (set A, Parody solution) and a set of parameters (B) which would correspond to an angular rotation rate *five times higher* than for A, i.e. $Le^2=2.5\ 10^{-3}$, $E=5\ 10^{-5}$, $E_m=5\ 10^{-2}$, and keeping the same background state (Fig. 13). When the forcing is increased, it appears that the attenuation becomes less and less negative while the mode frequencies are only slightly shifted. Thus, for the less attenuated modes, there exists a critical value of parameters where the linear solution exhibits a positive attenuation rate. The actual solution then becomes unstable and should evolve toward a saturated state or destabilize the mean flow. It can be mentioned that a few solutions behave differently, showing a non-monotonic path in the ($\lambda, \omega$) plane, as e.g. for $\omega$ = 1.33, or a path in the opposite direction, as for the only solution (set A) where $\lambda$ is positive ($\omega$ =4.72).

*m* = 3 results

For $m = 3$, the dominant experimental MC-mode is $S^a_{3-+}$ (G3 symmetry), its normalized frequency is $f_{exp}/\Delta f \approx 1.21$. The calculated eigensolutions show a behaviour similar to those calculated for $m = 1$ or 2 (not shown). Among these solutions, the modes with $\omega_{calc}/\Delta\Omega^* = 1.23$ (M1) and 1.33 (M2) present some similarities with the experimental $S^a_{3-+}$ mode (Fig. 15). In particular, the magnetic potential immediately outside the outer sphere is always localized below ~30° latitude. M1 mode exhibits the correct frequency but its magnetic potential is too much spiralized; the magnetic potential amplitude of the M2 mode follows well the experimental one, but its frequency is slightly too high and the spiralization too strong. Another characteristic is that regions near the inner sphere where the velocity exhibits a strong shear seem to follow here the background dipolar magnetic field lines, while the shear zones near the pole axis and close to the outer sphere appear comparatively less intense than for $m = 1$ or 2.

**IV. DISCUSSION**

The power spectra of the magnetic field fluctuations measured at the surface of the DTS experiment reveal a rich dynamical behavior. Our previous investigations [8] showed that the spectra of various observables (electric potential, magnetic field) were dominated by wide bumps, each bump corresponding to a single azimuthal wave number $m$. Here, we present measurements of the induced magnetic field over a complete meridian, in the case where the outer sphere is kept at rest. We can thus separate magnetic fluctuations that are symmetric with respect to the equator (Fig. 2a, 5b, 6a) from those that are anti-symmetric (Fig. 2b, 5a, 6b). Both symmetries are present, the former being dominant for $m=1$ and 3, while the latter dominates for $m=2$ (see Table 1). Due to the anti-symmetric nature of the applied magnetic field, the symmetries are reversed when the velocity field is considered. By correlating the signals measured at different latitudes, we find that several modes exhibit a clear spiralization (Fig. 4). For each azimuthal wave number $m$, we could reconstruct the scalar magnetic potential of the dominant mode at the surface (Fig. 8), by inversion of the correlations between the three components of the magnetic field. These new data synthesize best the signature of the various modes at the surface.

In order to grasp the physical nature of the modes, we have computed the solutions of the linearized MHD wave equations in a spherical shell, following the approach described in Ref. [9]. The waves develop over an imposed realistic background state given by the axisymmetric part of a 3D numerical simulation of the magnetized spherical Couette flow (Fig. 9). The magneto-inertial modes thus computed take into account the strong mean and differential rotations of the background state, and its imposed and induced magnetic field. The numerical results show that the imposed dipolar magnetic field profoundly modifies the distribution of the modes in the frequency-attenuation plane (Fig. 10), leaving only a limited set of modes, which display large-scale structures. For each azimuthal wave number $m=1$, 2 and 3, we have been able to find numerical modes that reproduce the signature of the measured magnetic potential (Fig. 11 and 12). The computed frequencies of the modes do not always match perfectly the measured frequencies. We think that this is due to discrepancies between the modeled and actual background states, which appear to greatly affect the frequency and attenuation of the modes (Fig. 10 and 13). This opens the interesting possibility of using the mode measurements to constrain the experimental mean axisymmetric flow. However, the mode eigenvalues do not evolve smoothly with the background state, making an inversion procedure difficult. We note that the best fitting modes are in a frequency range which is *not* associated with the weakest attenuation. This apparent paradox indicates that only these modes are efficiently excited in the DTS experiment, thus placing strong constraints on the excitation mechanism, as discussed below.

Although waves and instabilities are known to play a major role in geophysical and astrophysical flows, only few experiments relate to magneto-inertial waves. Non-axisymmetric patterns in a magnetized rotating spherical Couette flow have been taken as evidence of the magneto-rotational instability (MRI) [18]. An $m=1$ mode has been observed in the PROMISE Taylor-Couette experiment [19], in which a helicoidal MRI has also been detected. In another liquid-metal Taylor-Couette experiment, the observation of two $m=1$ modes with different frequencies has been interpreted as evidence for fast and slow magneto-Coriolis waves [20]. A richer set of sharp modes has been mapped in the Maryland rotating spherical Couette experiment [21]. There, the applied magnetic field was weak and the induced magnetic field could be used as a passive tracer of the flow. The surface signature and frequency of the modes were found to closely match those of the inertial modes of a full sphere. Only modes with an equatorially-antisymmetric flow were observed.

Since our numerical simulations provide a good match of the structures of the modes we measured in the DTS experiment, we can dwell further into their dynamics by examining the force balance. We plot the average intensity of the various forces in a meridional plane for the $m=2$ (Fig. 16) and $m=3$ (Fig. 17) dominant modes. The average intensity is computed as the square root of the squared force integrated over azimuth, i.e.:

$$<\mathbf{F}(r,\theta)> = \sqrt{\int \mathbf{F}^2 r \sin(\theta) d\phi} \qquad (5)$$

Three forces are considered: the inertial force (a), the Lorentz force (b), and the viscous force (c). The acceleration term is also plotted (d). The largest intensities are found in the inertial force (Fig. 16a and Fig. 17a), given by the $(\mathbf{U_0} \cdot \nabla)\mathbf{u} + (\mathbf{u} \cdot \nabla)\mathbf{U_0}$ term of Eq. 4. This term is dominated by the Coriolis force linked with the average solid body rotation of the fluid. The inertial force appears to balance most of the acceleration term (Fig. 16d and Fig. 17d), which is strong only in the outer region of the sphere. However, the Lorentz force (Fig. 16b and Fig. 17b) reaches comparable amplitudes and is the dominant force near the inner sphere, where the isovalues of its intensity appear to partly follow field lines of the imposed dipolar field. Viscous forces (Fig. 16c and Fig. 17c) only contribute at the outer boundary. We also plot the kinetic energy density (Fig 16e and Fig. 17e) and the magnetic energy density (Fig 16f and Fig. 17f) of the modes. The total kinetic energy is between two and three orders of magnitude larger than the total magnetic energy! In our regime with large magnetic diffusivity and a strong imposed magnetic field, we expect magnetic field fluctuations to scale as $R_m B_0$. Then, the magnetic over kinetic energy ratio scales as $Lu_2$, where $Lu = aB_0/\eta\sqrt{\rho\mu} = LeR_m$ is the Lundquist number, which has a value of 1 in our numerics, twice the experimental value. Therefore, in both cases, we could have expected the kinetic and magnetic energies to be comparable. It is interesting to note that magnetic fluctuations reach comparable amplitudes throughout the fluid shell although velocity fluctuations are very small near the inner sphere. This is of course a consequence of the intensity distribution of the imposed magnetic field, due to its dipolar nature.

The magneto-inertial modes that we observe and model are thus characterized by a strong influence of the Lorentz force, which inhibits fluctuations where the imposed magnetic field is largest. The region of influence of the magnetic field roughly covers the zone where the local Elsasser number $\Lambda(r) = \sigma B^2(r)/\rho\Omega$ is larger than one, as found by Brito et al [7] for the DTS mean flow. However, the strong magnetic diffusion, measured by the small value of the Lundquist number, present in DTS and other liquid metal experiments, inhibits the propagation of Alfvén waves. Therefore, even where it is strong, the Lorentz force is not in balance with the acceleration term, but rather with the pressure gradient. Another indication of the non-Alfvénic character of our magneto-inertial modes is the small ratio of magnetic over kinetic energy. It is important to keep this limitation in mind when trying to extrapolate experimental results to geophysical and astrophysical situations, for which magnetic diffusion is negligible. In that respect, notions such as fast and slow waves [20], or Alfvénic MRI, are not relevant in the experimental context. Our analysis is in line with the conclusions of Gissinger et al [22] and Roach et al. [23], who relate the modes observed in the Maryland and Princeton liquid metal Couette flows to MHD instabilities arising from internal or surface shear layers.

Finally, we wish to discuss the excitation mechanism of the modes we observe. Several different explanations have been put forward to explain the excitation of the Maryland inertial modes: over-reflection off the Stewartson layer [24], or instabilities of the inner sphere boundary layer inside the tangent cylinder. In both cases, it remains unclear why only equatorially anti-symmetric modes are excited. Instabilities of magnetically induced shear layers have been also invoked in numerical studies of spherical Couette flows [25, 26]. In the DTS geometry, one can think of two regions prone to exciting modes: the external boundary layer, where strong shear injects energy, and the zone of super-rotation near the inner sphere, where small fluctuations produce large Lorentz forces. The meridional maps of kinetic energy favor the first hypothesis. In recent numerical 3D simulations of the magnetized spherical Couette flow, it has been found that bumps naturally appear in spectra of long time-series of the surface magnetic field (also see Matsui et al [27]). These bumps share the properties of the modes observed in DTS and can be linked to instabilities of the external shear layer. This layer is of the Bödewadt type, which is known to be particularly unstable [28]. This excitation mechanism could explain the paradox of the strong attenuation of the best fitting modes, as viscous friction in the boundary layer is the dominant dissipation mechanism of the modes.

**ACKNOWLEDGMENTS**

N. Gillet, D. Jault and N. Schaeffer are gratefully acknowledged for stimulating discussions.

**FIGURE**

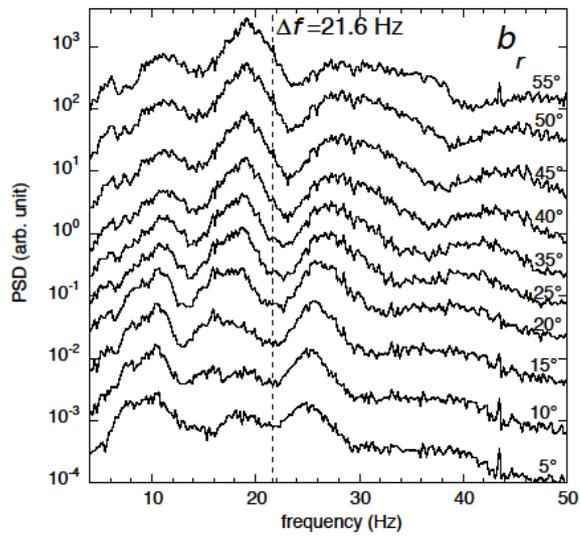

FIG. 1. Power spectral density of total radial magnetic component $b_r$, at various latitudes ranging from +5° (bottom) to +55° (top), as indicated; successive curves are shifted upwards for clarity; dashed line shows the forcing rate $\Delta f = 21.6$ Hz.

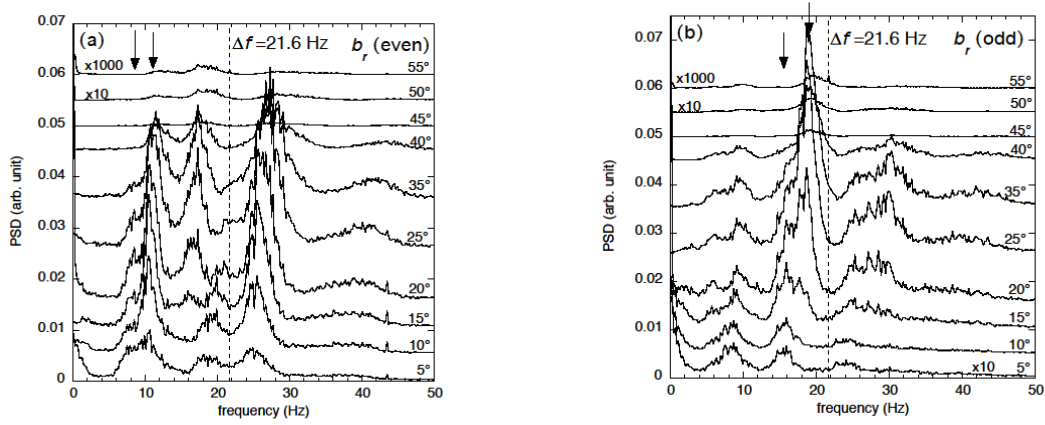

FIG. 2. Symmetrized power spectral densities of radial fluctuating magnetic field, at various latitudes ranging from 5° (bottom) to 55° (top): (a) even part, (b) odd part; arrows indicate the frequency of sub-components for $m = 1$ (a) and $m = 2$ (b); the weakest signals are multiplied by a factor as indicated, and each curve is shifted vertically by a constant value, for sake of clarity; dashed line shows the forcing rate $\Delta f = 21.6$ Hz.

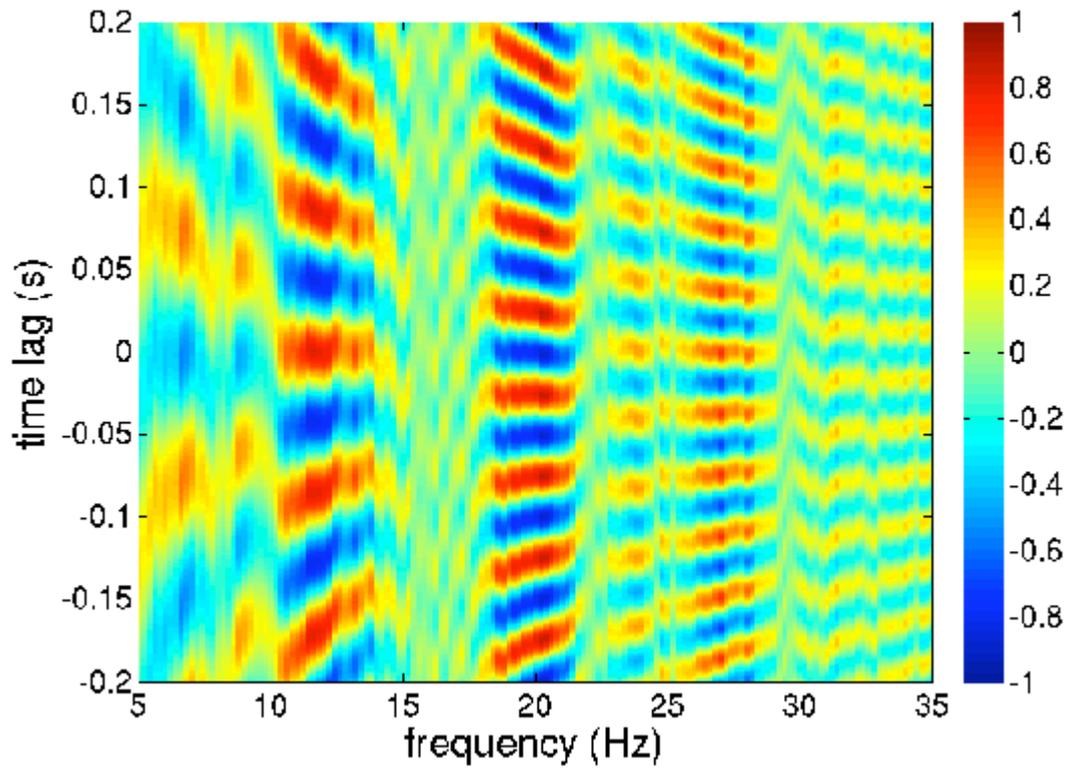

FIG. 3. Correlation time - frequency diagram between pair of radial component $b_r$ located at ±35° latitude for $\Delta f$ = 21.6 Hz; correlations are normalized (see text); at zero lag, red (blue) spots mean correlated (anti-correlated) data.

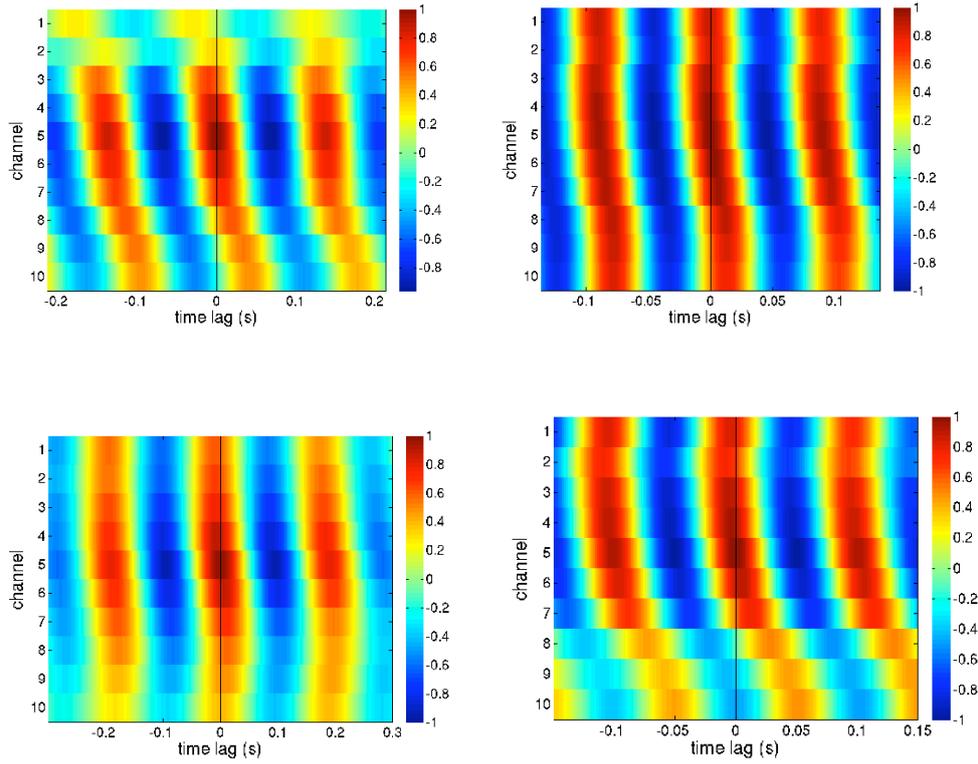

FIG.4. Latitudinal dependence of the correlation between the radial magnetic field at 35° latitude (channel 5) and the other latitudes ($m = 1$); (a, b): even parity, (c, d): odd parity; frequency windows are [6-8] Hz (a), [10-12] Hz (b), [4-6] Hz (c), [9-11] Hz (d); equator is at bottom of each figure; note the high spiralization for (a) and (d) modes.

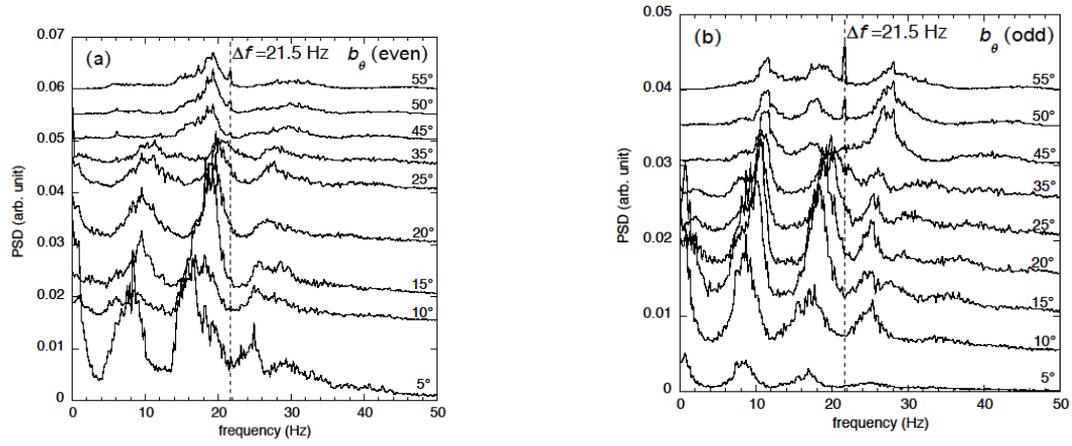

FIG.5. Symmetrized power spectral densities of meridional fluctuating magnetic field, at various latitudes ranging from 5° (bottom) to 55° (top): (a) even part, (b) odd part; for sake of clarity, each successive curve is shifted vertically; dashed line shows the forcing rate $\Delta f = 21.5$ Hz.

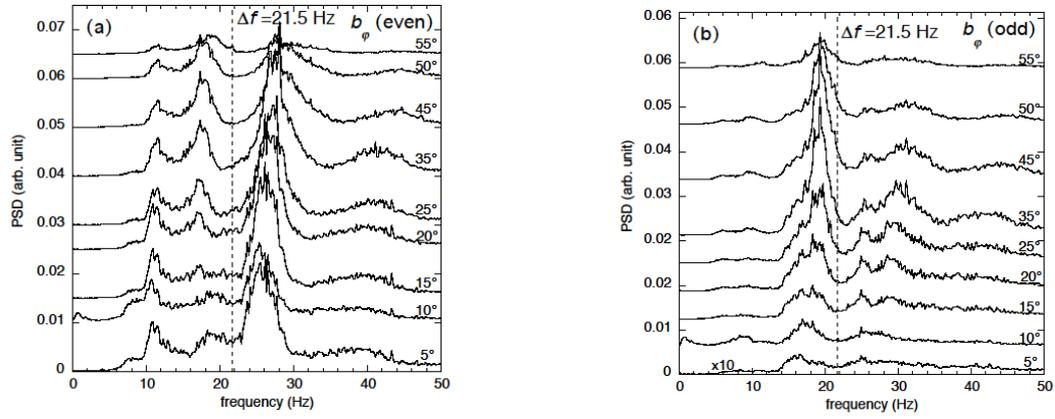

FIG.6. Symmetrized power spectral densities of azimuthal fluctuating magnetic field, at various latitudes ranging from 5° (bottom) to 55° (top): (a) even part, (b) odd part; for sake of clarity, each successive curve is shifted vertically; dashed line shows the forcing rate $\Delta f$ = 21.5 Hz.

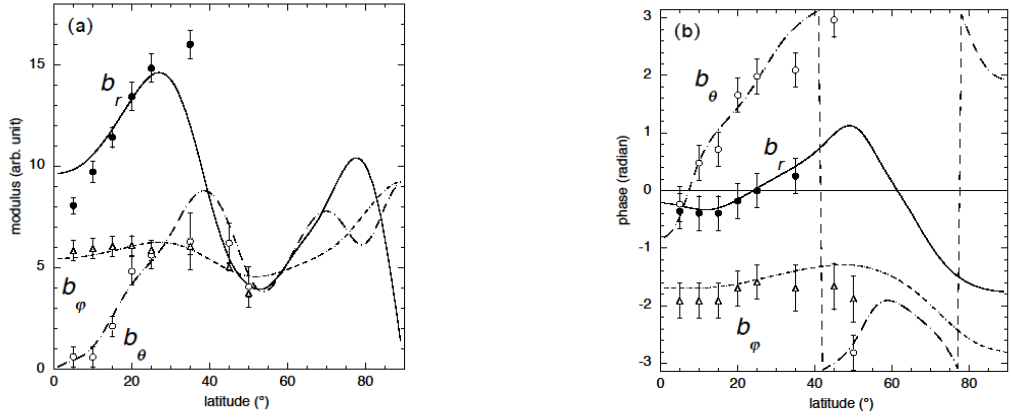

FIG.7. Latitudinal variation of the modulus (a) and phase (b) of the complex magnetic field components for the mode $S^b_{1-+}$; symbols are extracted from cross-correlation data; lines are calculated; $b_r$: full circles, continuous lines; $b_\theta$: open circles, dashed lines; $b_\varphi$: triangles, dotted lines.

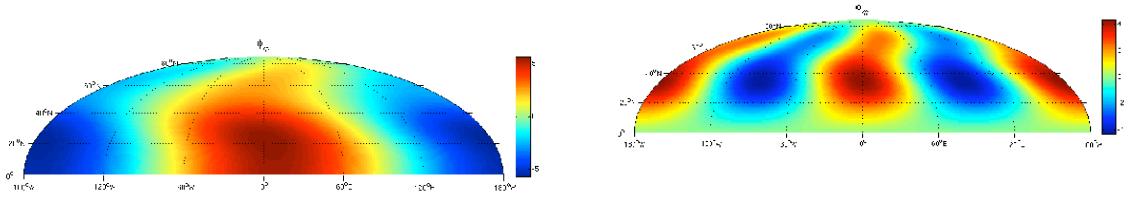

FIG.8. Magnetic potential reconstructed at the surface of the outer sphere for the $S^b_{1-+}$ (a) and $S^b_{2+-}$ (b) modes, shown as a Mollweide projection.

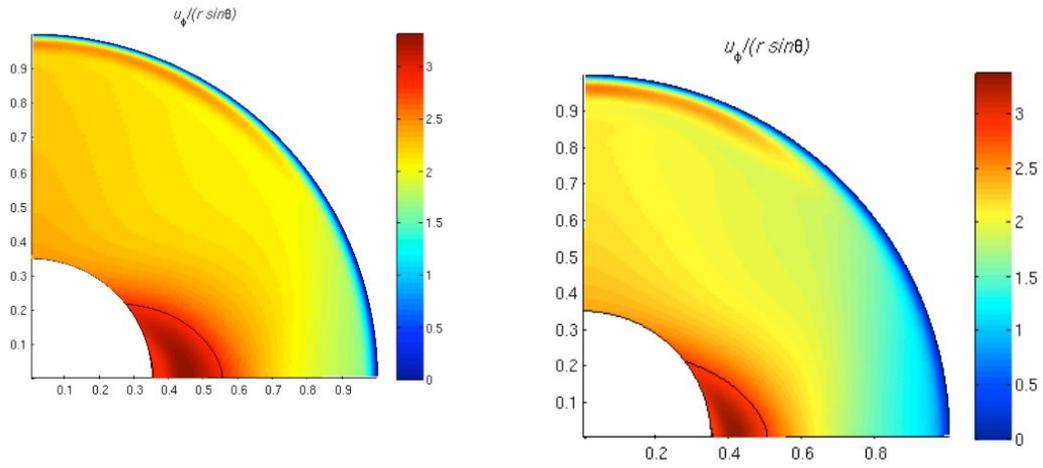

FIG.9. a) Meridional dependence of angular velocity $u_\phi/(r \sin\theta)$ for the background field $\mathbf{U_0}$, from (a) PARODY and (b) XSHELLS codes; note the superrotation region where the fluid rotation rate is higher than the dimensionless inner sphere rotation rate $\Delta\Omega^* = 2.86$ (black line corresponds to this value).

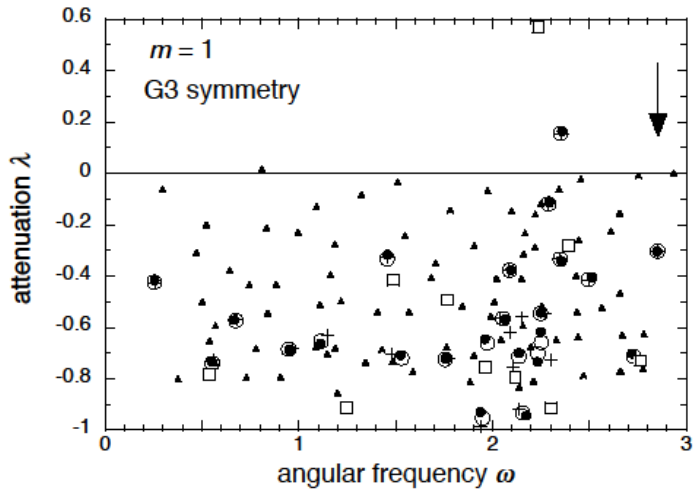

FIG.10. Attenuation $\lambda$ versus angular frequency wcalc for $m = 1$ modes (G3 symmetry): PARODY background state with dipolar field alone and conducting inner sphere (circles), insulating inner sphere (+) or $Le = 0$ (triangles); same background state but with induced field added (full circles); solutions with XSHELLS background state (squares, no induced field, conducting inner sphere); the arrow shows the inner sphere rotation rate $\Delta\Omega^* = 2.86$.

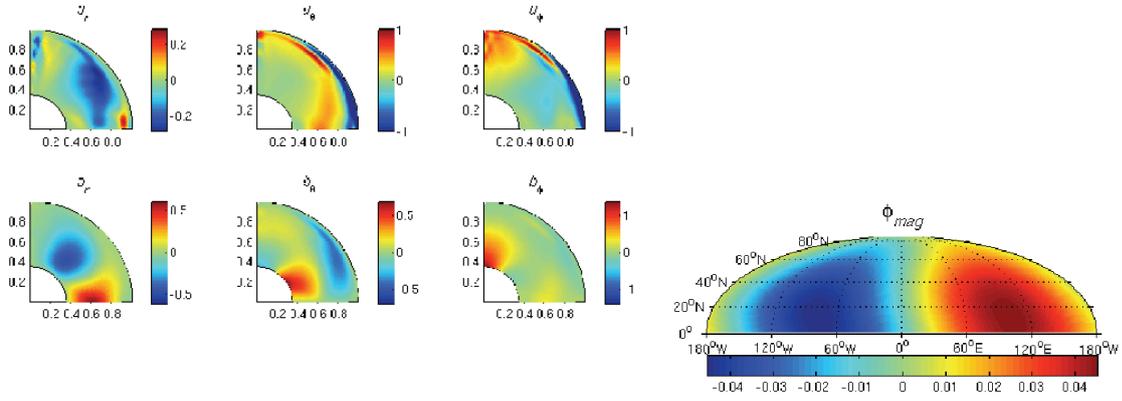

FIG.11. (a) Meridional sections ($\varphi = 10°$) of the three spherical components of **u** and **b** for $m = 1$ MC-mode with angular frequency $\omega_{calc}/\Delta\Omega^* = 0.39$ (G3 symmetry); (b) associated magnetic potential at the outer sphere surface, shown as a Mollweide projection.

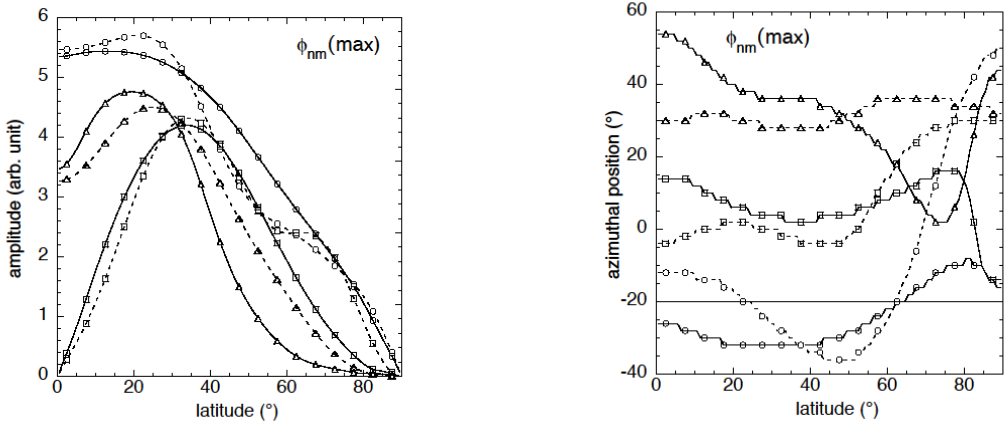

FIG.12. Latitudinal variation of the amplitude (a) and azimuthal position (b) of the maximum of magnetic potential for the three most intense modes: $S^b_{1-+}$ (circles), $S^b_{2+-}$ (squares) and $S^a_{3-+}$ (triangles); continuous lines are calculated, dashed lines are guides for the eyes for experimental data.

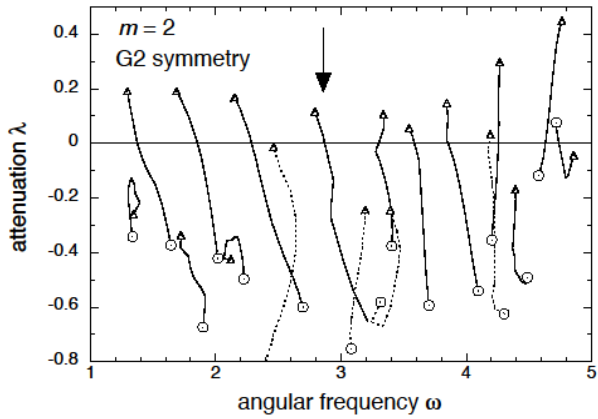

FIG.13. Attenuation $\lambda$ versus angular frequency $\omega_{calc}$ for $m = 2$ modes (G2 symmetry): PARODY background state with dipolar field alone and conducting inner sphere (circles, set of parameters A) ; the arrow shows the inner sphere rotation rate $\Delta\Omega^* = 2.86$; triangles are solutions for the same background state but for dimensionless parameters corresponding to a rotation rate five times higher (set B, see text); continuous and dotted lines indicate how the eigensolutions evolve between A and B.

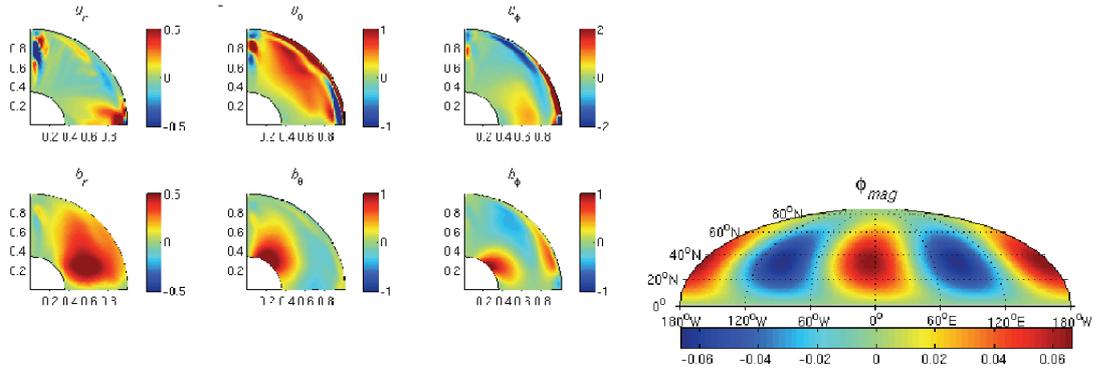

FIG.14. (a) Meridional sections ($\varphi = 10°$) of the three spherical components of **u** and **b** for $m = 2$ MC-mode with angular frequency $\omega_{calc}/\Delta\Omega^* = 0.94$ (G2 symmetry); (b) associated magnetic potential at the outer sphere surface, shown as a Mollweide projection.

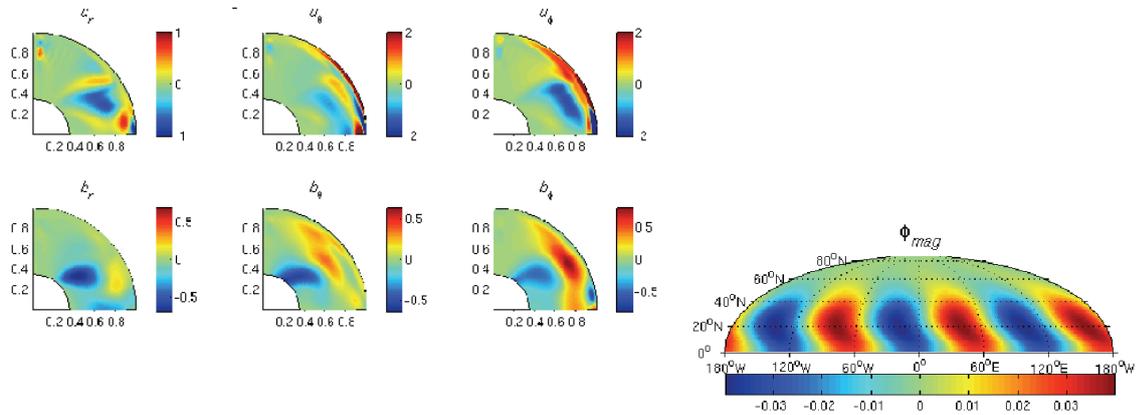

FIG.15. (a) Meridional sections ($\varphi = 10°$) of the three spherical components of **u** and **b** for $m = 3$ MC-mode with angular frequency $\omega_{calc}/\Delta\Omega^* = 1.33$ (G3 symmetry); (b) associated magnetic potential at the outer sphere surface, shown as a Mollweide projection.

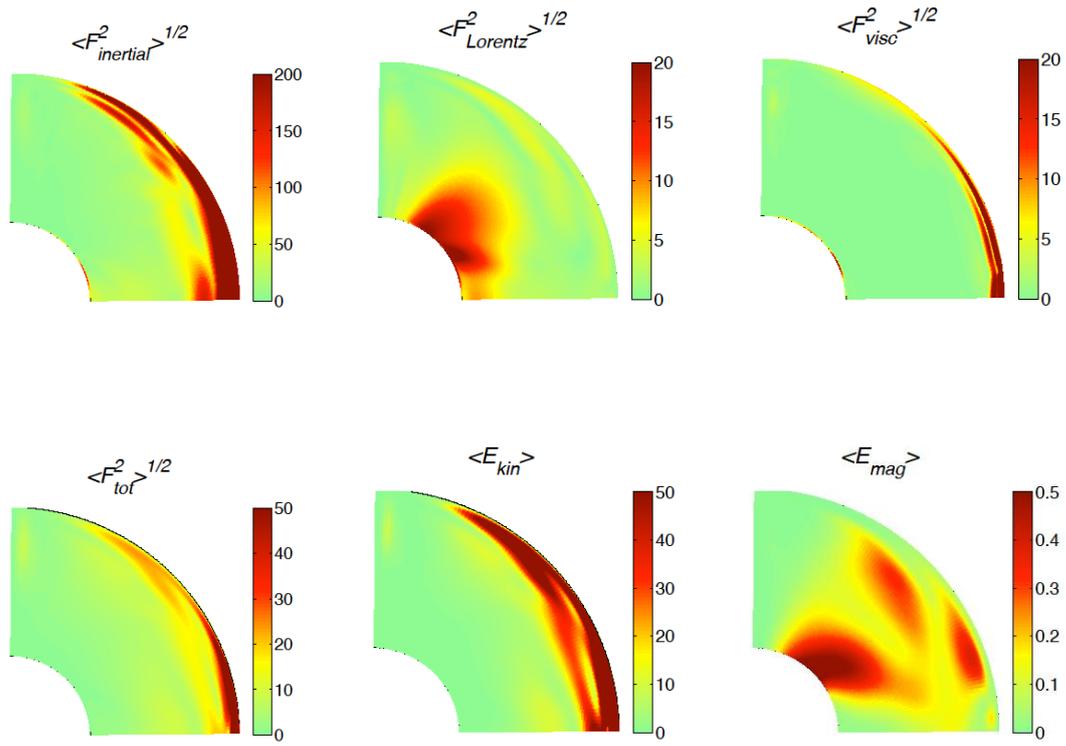

FIG.16. Meridional sections of: (a) inertial, (b) Lorentz, (c) viscous and (d) total forces, for the $m = 2$ MC-mode shown in Fig. 14; kinetic (e) and magnetic (f) energy density for the same mode; all values are non-dimensioned and integrated along the azimuthal direction.

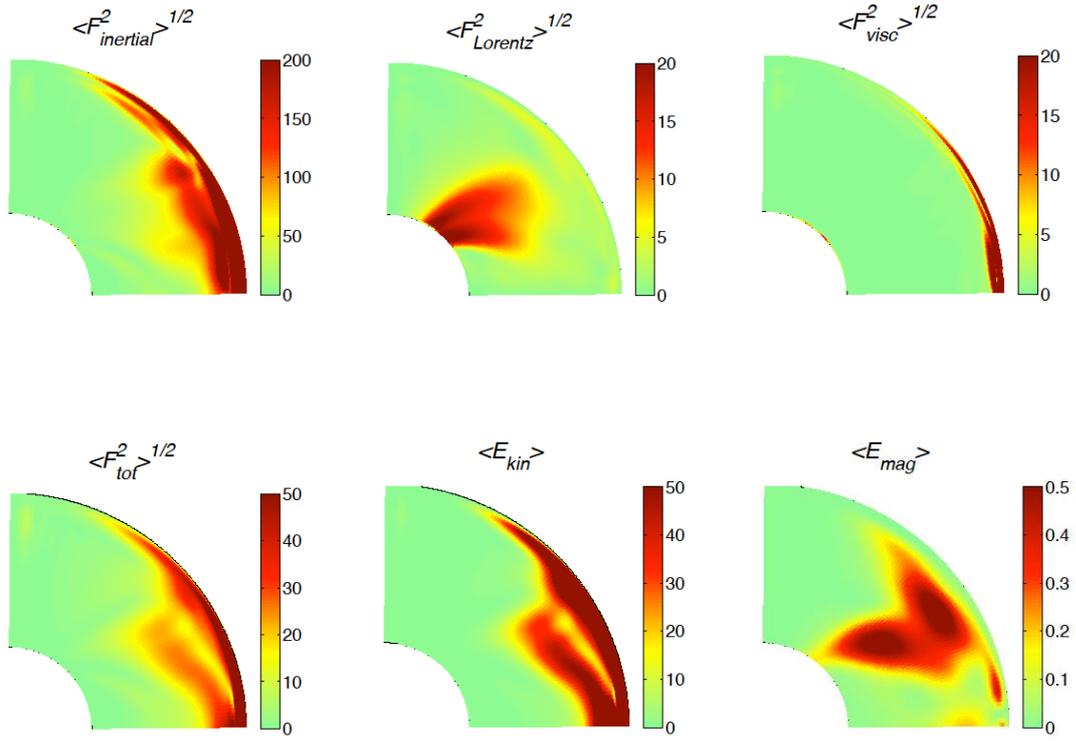

FIG.17. Meridional sections of: (a) inertial, (b) Lorentz, (c) viscous and (d) total forces, for the $m = 3$ MC-mode shown in Fig. 15; kinetic (e) and magnetic (f) energy density for the same mode; all values are non-dimensioned and integrated along the azimuthal direction.

## Table

Frequency $f_{exp}$ of the MC-modes identified in DTS experiment for both G2 and G3 symmetries ($\Delta f = 21.5$ Hz); the three dominant modes are shown in bold.

|  |  | $m = 1$ | | $m = 2$ | | $m = 3$ | | $m = 4$ |
|---|---|---|---|---|---|---|---|---|
| G2 symmetry | mode | $S^a_{1+-}$ | $S^b_{1+-}$ | $S^a_{2+-}$ | $S^b_{2+-}$ | $S^a_{3+-}$ | $S^b_{3+-}$ | $S_{4+-}$ |
|  | $f_{exp}$ (Hz) | 6 | 9 | 15.5 | **19** | 25 | 30 | 42 |
| G3 symmetry | mode | $S^a_{1-+}$ | $S^b_{1-+}$ | $S^a_{2-+}$ | $S^b_{2-+}$ | $S^a_{3-+}$ | $S^b_{3-+}$ | $S_{4-+}$ |
|  | $f_{exp}$ (Hz) | 8.5 | **11** | 16.5 | 20 | **26** | 30.5 | 41 |